\documentclass[sigconf]{acmart}
\AtBeginDocument{%
  }

\setcopyright{acmlicensed}
\copyrightyear{2016}
\acmYear{2016}
\acmDOI{...}
\acmConference[AUI'16]{Asian Workshop on User Interface}{June 03--05, 2016}{Tokyo, Japan}

\begin{document}

\title{Flash Word : ESL Listening Support with Subtitle Skimming} 

\begin{abstract}
Listening to English audio with English subtitles is thought to be effective for ESL students to improve listening comprehension skills. However, for those who learn English as a second language it can be hard to finish English subtitles. Also they are unlikely to focus on listening while being busy reading subtitles. 
We propose \textit{Flash Word}, a subtitling system which generates and presents shortened intralingual subtitles for English audio by detecting keywords from the script. Flash Word extracts keywords from scripts using an existing algorithm, and shows them one at a time like a flashcard. The system calculates when the keywords are being pronounced and aligns them correctly. 
\textit{Flash Word} aims to enable ESL students to focus better on listening by minimising the cost of reading subtitles.
\end{abstract}

\author{Yuzu Saijo}
\affiliation{%
\institution{The University of Tokyo}
\city{Tokyo} 
\country{Japan} 
}

\author{Jun Rekimoto}
  \affiliation{%
  \institution{The University of Tokyo}
  \city{Tokyo}
  \country{Japan}}
\affiliation{%
  \institution{Sony Computer Science Laboratories}
  \city{Tokyo}
  \country{Japan}}


\begin{teaserfigure}
\centering
  \includegraphics[width=0.9\linewidth]{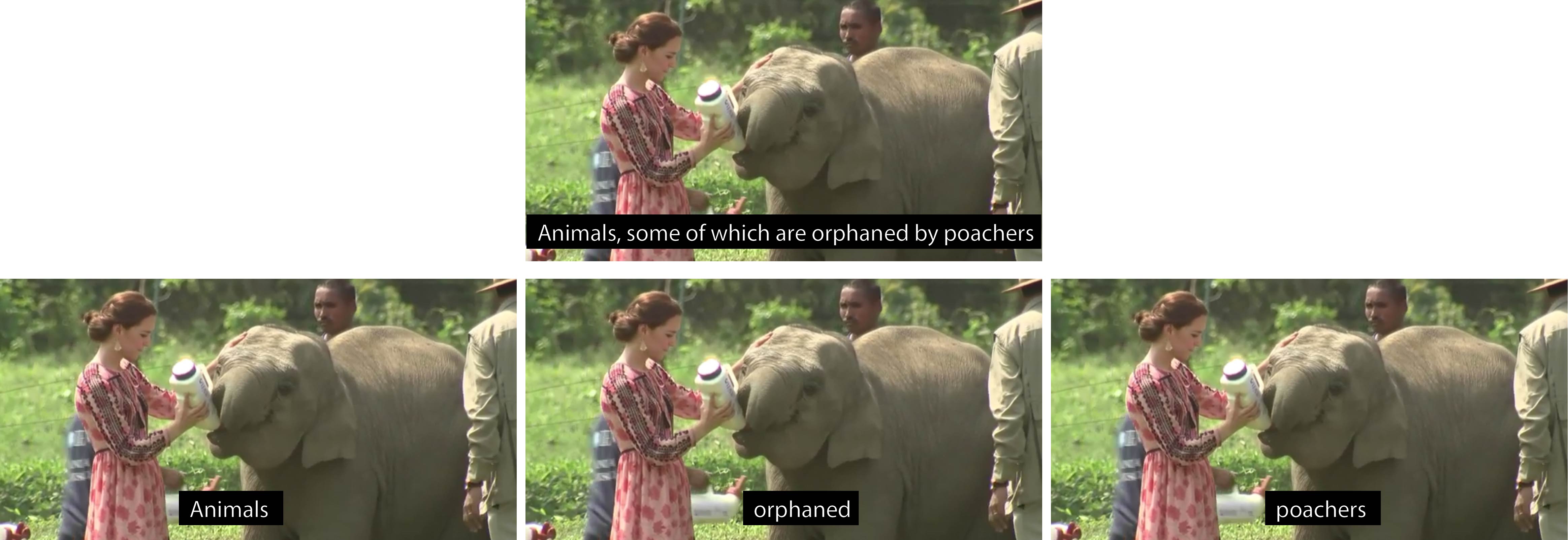}
  \caption{A comparison of a subtitle with and without skimming (video source\cite{bbc}). (above) Normal subtitling. (below) Our proposed method that only displays selected words from the script, synchronized with the audio, letting ESL students concentrate on listening, rather than reading subtitles. }
  \label{fig:comp}
\end{teaserfigure}

\maketitle


\keywords{
ESL, Language Learning, Listening, Subtitling, Skimming
}

\section{Introduction}

The English language is spoken globally today and there is a growing need for effective ways to study English as a second language.
In learning English as a second language (ESL), listening comprehension skill is crucial among others. 
In fact, out of the time spent in English communication, it is said that more than half of the total time is spent on listening~\cite{abbas}.
One of the methods that are proved to improve listening comprehension is to listen to English audio with English subtitles~\cite{antar08,field09}.

However, the speed of reading subtitles differs from individual to individual, and ESL students are assumed to be slower in reading English than native speakers. 
Though professional subtitle writers often omit unnecessary words from original scripts to shorten subtitles~\cite{caimi06}, ESL students are still likely not to be able to finish the English subtitle of the audio they listen to. Thus, there is a possibility where a student pay too much attentions to subtitles, neglecting listening concentration.

Based on this observation, we would like to propose a hypothesis that presenting only important words should be more helpful for ESL learners to better understand English audio than showing the whole script. 
We also consider that this approach is valuable in terms of human augmentation. Instead of spoiling the listening skills by full subtitles, this approach rather supplements or augments the user’s listening ability with carefully selected keywords.
In order to test and support this hypothesis, we have designed a system called \textit{Flash Word}, which generates a shortened subtitle by detecting keywords in the English audio.

In this paper, we describe the idea of subtitle skimming as listening support, proposing two possibilities of the system design, and explain the system architecture of the prototype of \textit{Flash Word}. Then we describe the pilot study using the present prototype, analyze its results and finally we discuss its future work.

\section{Listening Support with Skimmed Subtitling}

Figure~\ref{fig:comp} shows the comparison of typical English listening training titles with the full subtitle (above) and the skimmed one (below).  As shown in this figure, the full subtitle version might pose to much attention to the listeners for reading the whole sentence, while the skimmed version the listener can concentrate on audio and glancing the keywords.  These skimmed words synchronously appear on the screen with the audio, so that the listener can promptly know the important word even when the listener missed it.

\textit{Flash Word} can be designed either for pre-subtitling with scripts or for real-time subtitling without scripts. 

\begin{figure}
    \centering
    \includegraphics[width=0.8\linewidth]{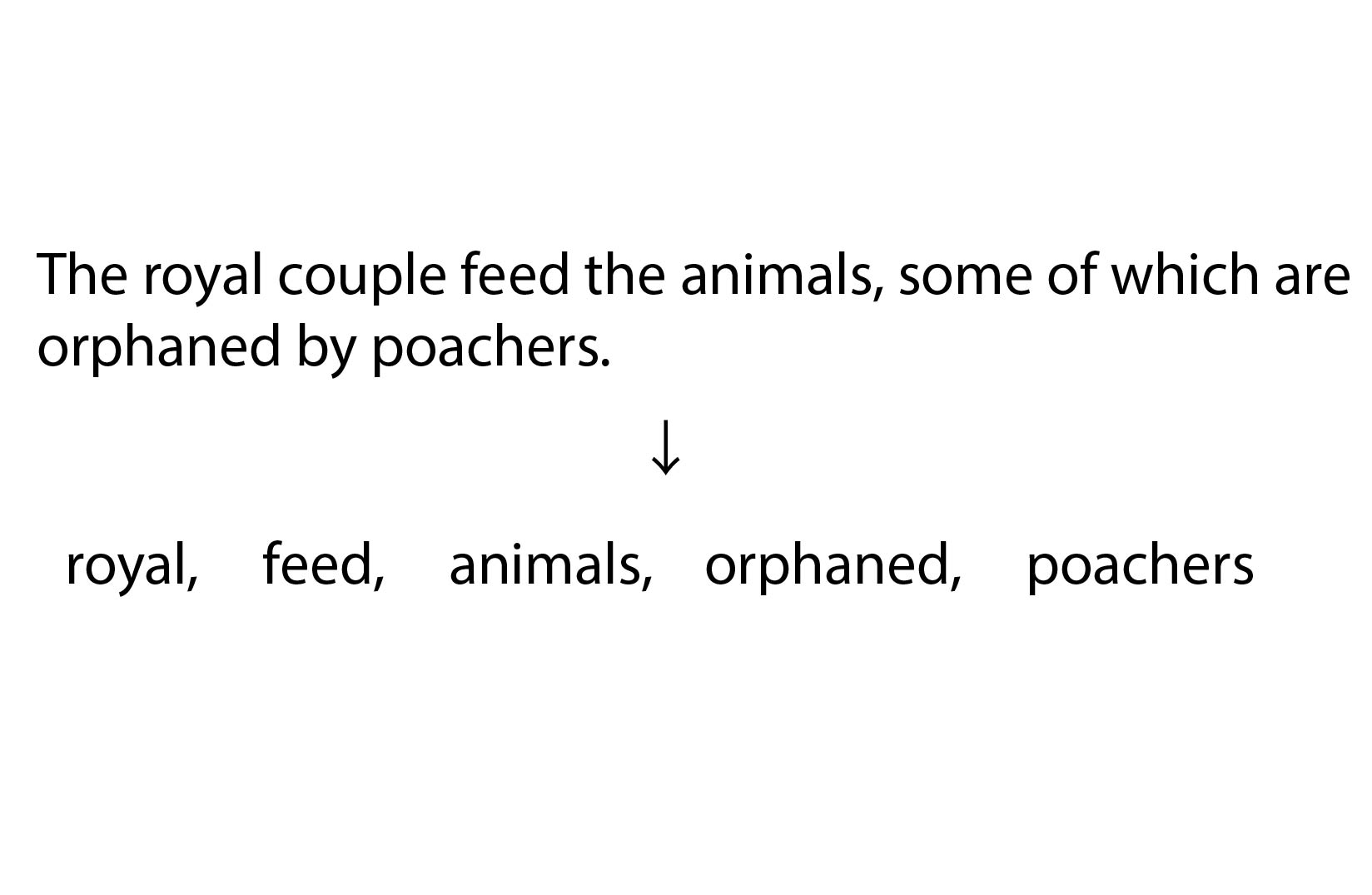}
    \caption{An example of subtitle skimming with RAKE algorithm}
    \label{fig:shortened}
\end{figure}

\subsection{Pre-subtitling with Scripts}
When an English audio has its script, \textit{Flash Word} prepares a shortened subtitle beforehand.
In order to shorten the subtitle, the present system plans to utilize three methods of detecting important words in the context;
 it quantifies each word's importance (1) by tf-idf (term frequency - inverse document frequency) formula~\cite{Jones72astatistical}, (2) by Rapid Automatic Keyword Extraction (RAKE) algorithm \cite{rake}, which detects key phrases based on registered stop-words, and (3) by using TextRank\cite{textRank}, an algorithm which quantifies each word's significance based on the context.

To correctly synchronize speech with the subtitle, the system also provides an alignment mechanism for detecting word pronunciation timing based on the given audio and its corresponding script.  To do this, we use a technique developed for our previous system~\cite{withyou}. The system first applies a speech recognition over the audio file using Julius~\cite{Lee01julius}.  As the Julius system provides a time point of each recognized word, this information can be used for the alignment.  Then, the compares the script with recognized text and correct miss-recognitions.  Finally, we get the correct time point of each word in the script.

\section{System Architecture}
\begin{figure*}[htbp]
\begin{center}
\includegraphics[width=0.9\linewidth]{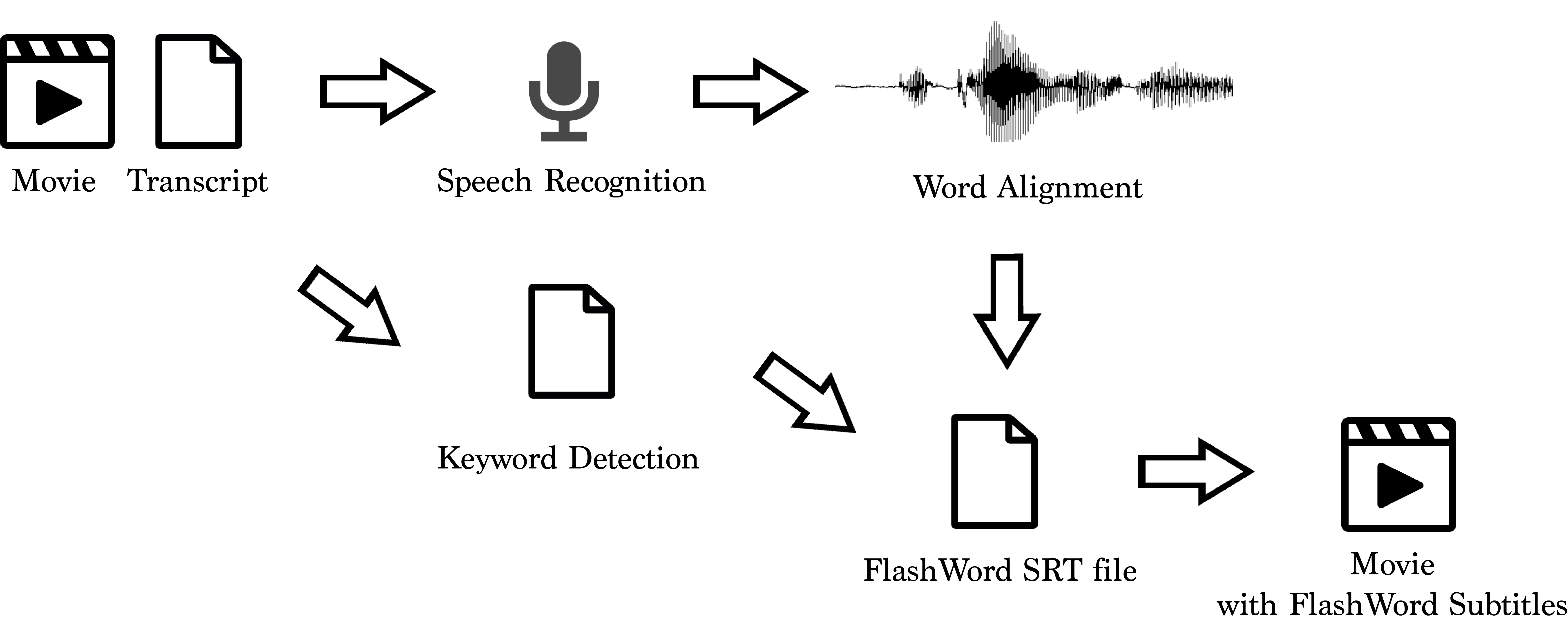}
\caption{Flash Word takes a movie which includes speech in English and its transcript. It aligns the words in the script to the movie by speech recognition. Then the system combines the alignment and the detected keywords and generates a subtitle file for the movie. }
\label{fig:sys}
\end{center}
\end{figure*}

We have developed a pre-subtitling prototype of \textit{Flash Word}. In this section, we explain how the prototype is implemented. Figure~\ref{fig:sys} shows the whole system architecture. The pre-subtitling prototype of \textit{Flash Word} takes a movie and its transcript as inputs and returns a \textit{Flash Word} subtitle file for the movie. The prototype is designed for use in experiments, so that it can work best for relatively short movies such as news stories, not for long ones such as films.

\subsection{Speech recognition and Word alignment}
As described in the previous section, in order to adjust the timing of flashing subtitles to audio, the present system utilizes Julius\cite{Lee01julius}, which is an open-source large vocabulary continuous speech recognition engine. Julius needs an acoustic model which links acoustic features and linguistic features, and a language model which decides the probability that a certain word string appears in the language in order to recognize speech correctly\cite{withyou}. In our prototype, we employ a free English acoustic model provided online for an acoustic model. For a language model, the prototype system generates an English language model based on the transcript of the audio. That is, the system takes the transcript sentences as a grammar pattern which can be uttered. 
\textit{Flash Word} needs to detect when the keywords are pronounced in the given audio. So as to realize this, the prototype system operates Forced Alignment in Julius, which calculates when in the audio each word from the transcript is being spoken. By Forced Alignment, the system generates a list of all the words from the transcript and their timings of pronunciation.

\subsection{Keyword detection}
As mentioned above, there are several possibilities of how to detect keywords in the English speech. We tried extracting keywords using Tf-idf, RAKE and TextRank. Out of those three methods, RAKE algorithm turned out the best for short scripts. Both Tf-idf and TextRank quantify word importance based heavily on its occurrence frequency. Since not many words appear multiple times in short scripts, neither of them extracted important keywords. Tf-idf with the short scripts resulted in most of the words in the scripts having the same score, and TextRank was unable to generate a sufficient graph.
For example, the sentence shown in Figure\ref{fig:shortened}, which is "The royal couple feed the animals, some of which are orphaned by poachers" results in no keywords by tf-idf calculation, and in only "royal couple" by TextRank.

Thus, the prototype utilizes only RAKE algorithm to detect keywords from a script. RAKE algorithm recognizes phrases by separating a script with pre-registered stop-words, such as prepositions, articles and basic adjectives\cite{rake}. And it scores phrases based on the frequency of words in a phrase and length of a phrase\cite{rake}. 

\subsection{Subtitle generation}
After aligning the script with the audio and detecting the keywords, by combining the results the system generates a subtitle file in an SRT format, which can be integrated into videos as subtitles easily on different platforms. The system runs a ruby script which iterates through the list of words and pronunciation timings and generates a new list that only contains keywords and their pronunciation timings. In this process the timings of the keywords pronunciation are adjusted so that a keyword is being shown as a subtitle until the next keyword appears. The system converts the generated list into an SRT format, and integrates it into the input video.

\section{Pilot Study}
We conducted a pilot study in order to test if \textit{Flash Word} actually enables ESL students to focus on audio better, and also to collect feedback from participants for the purpose of future implementation.
\subsection{Experimental design}
We recruited 12 participants, aged 22-25 years old (average: 23.6 years, SD: 0.96 years), nine male and four female, from the local university. The participants had no prior association with our laboratory or with our project. Students' majors included physics, architecture and art. No participants were from English speaking countries and all the participants identified themselves as ESL students.

The pilot study consisted of four parts:  (1) pre-test questionnaire, (2) introduction of the system, (3) three listening comprehension tests, and (4) post-test questionnaire and interview.

Before conducting listening comprehension tests, we asked the participants to answer a demographic questionnaire. The questions also included TOEIC, TOEFL scores so that we could know the English skills.

For listening comprehension tests, we prepared three different listening audio from CNN Student News\cite{cnn}. Each audio is cut down to 19 seconds and contains (A)57, (B)54 and (C)52 words. Readability of each transcript measured on online readability calculator\cite{readability} is (A)6.8, (B)7.0 and (C)7.0. None of the audio contents were related to each other. 
From audio (A), (B) and (C), we prepared videos which have white blank background with each audio. For each video, we prepared three subtitle option: no subtitles, normal subtitles and \textit{Flash Word} subtitles.

We divided 12 participants into three groups: Group 1 watched (A) with no subtitles, (B) with normal subtitles, and (C) with \textit{Flash Word} subtitles. Group 2 watched (A) with normal subtitles, (B) with \textit{Flash Word} subtitles, and (C) with no subtitles. Group 3 watched (A) with \textit{Flash Word} subtitles, (B) with no subtitles, and (C) with normal subtitles. We changed the order in which each group sees three types of subtitles so that the order or the difference in difficulties of questions would not matter.

We asked the participants to watch videos three times each, and then answer three questions we prepared beforehand for each video. 
So, each participant watched video(A), (B) and (C) three times, and answered three questions each for (A), (B) and (C) after watching each of them.

After the tests, we asked the participants to answer post-test questionnaire. We also interviewed them and recorded their feedback.

\subsection{Subtitles for test videos}
We prepared three subtitle options for each (A), (B) and (C) video. Normal subtitles show the exact transcript, around 15 to 20 words at a time with the timings of pronunciation. \textit{Flash Word} subtitles show only the keywords extracted with RAKE algorithm. The numbers of the keywords shown for each video are (A)18, (B)23, and (C) 23.

\subsection{Listening comprehension tests}
One of the purposes of this pilot study was to test if \textit{Flash Word} enables participants to focus on audio than normal subtitles. In order to examine it, we prepared three questions for each video that asks about audio information. 
Table\ref{tab:lis} shows an example of the questions we actually asked. The questions were design to focus on how words are pronounced in the audio, so that a participant would not be able to answer them correctly if he or she reads subtitles and remember them without listening to the audio.
All the questions are multiple choice problems.

\begin{table}[htb]
\caption{Example of listening comprehension questions}
\label{tab:lis}
\centering
  \begin{tabular}{p{1.8cm} p{5cm}}
    Normal Subtitle & 0600 hours, NASA prepares to fly. \rule[0mm]{0mm}{5mm}\\ \hline
    Pronunciation & O[ou] six hundred hours, NASA prepares to fly.\rule[0mm]{0mm}{5mm} \\\hline\hline
    Question & How was the word "0600" pronounced in the video? (1) O[ou] six double O[ou] (2) Zero six hundred (3) O[ou] six hundred (4) Zero six double zero  \rule[0mm]{0mm}{5mm} \\\hline
  \end{tabular}
\end{table}

\subsection{Results of tests}
The result of the listening comprehension tests is shown in Table\ref{tab:score}. The result shows that in this pilot study, the participants performed best with no subtitles, followed by normal subtitles, and worst with \textit{Flash Word} subtitles.

\begin{table}[htb]
\caption{Result of listening comprehension test : numbers in the table show mean scores out of three points}
\label{tab:score}
\centering
  \begin{tabular}{p{2cm} p{2cm} p{2cm}}
    No subtitles & Normal subtitles & Flash Word Subtitles\rule[0mm]{0mm}{5mm}\\ \hline
    1.42 & 1.25 & 1.17\rule[0mm]{0mm}{5mm}\\ \hline
  \end{tabular}
\end{table}

\subsection{Feedback from questionnaire and interview}
We asked questions about \textit{Flash Word}  via questionnaire after the listening tests. The results are shown in Table\ref{tab:ques}, which shows the mean, SD, and percentage of positive responses (>4 on a 7-point Likert scale) to each question. 

\begin{table}[htb]
\caption{Result of Questionnaire}
\label{tab:ques}
\centering
  \begin{tabular}{p{3cm} p{1cm} p{1cm} p{1cm}}
    Question & Mean & SD & \% \\ \hline \hline
    Was it easy to listen to the audio with Flash Word? & 4.46 & 0.96 & 9/13\rule[0mm]{0mm}{5mm}  \\ \hline
    Do you want to recommend Flash Word to other ESL students? & 5.00 & 1.22 & 7/13\rule[0mm]{0mm}{5mm}  \\\hline
  \end{tabular}
\end{table}

Some of the comments from the post-test interview are show in Table\ref{tab:interview}. The table lists the participants' TOEIC English score next to their feedback.
\begin{table}[htb]
\caption{Feedback from Interview}
\label{tab:interview}
\centering
  \begin{tabular}{p{2cm} p{5cm}}
    Participant's TOEIC score & Feedback\\ \hline \hline
    400 & {\it "To be honest I'm not used to English listening so I just read subtitles when they were there.}   \\\hline
    550 & {\it "I understood the audio with Flash Word subtitles, but couldn't really focus on listening.}" \\\hline
    760 & {\it "It was easy to guess what was being said in the audio with Flash Word."}  \\\hline
    900 & {\it "I would want this Flash Word for studying English listening. I think ESL students who are already good at English would benefit the most from this."}   \\\hline
  \end{tabular}
\end{table}

\subsection{Analysis}
As a result of the pilot study, we were not able to show that the participants focus better on audio with \textit{Flash Word} than with normal subtitles. However, it is assumed from the interview that ESL students with relatively higher scores appreciate \textit{Flash Word} better. In this pilot study we do not have all the participants' TOEIC scores so it is hard to classify the test performance based on the English ability, but in future experiments we plan to expand the experiments to more participants so that we could have more data to show whether \textit{Flash Word} system works good for those who are at specific English levels.

\section{Discussion and future work}
We have developed a pre-subtitling prototype of \textit{Flash Word}.
One possible way of enhancing the proposed system is to determine keywords to be displayed in accordance with the student's vocabulary level.  By measuring it, the system would be able to show difficult or unknown words tailored for the particular student.  

Other than ESL students support, we also consider that this system would be beneficial to senior people who tend to miss the spoken words even in their own mother tongue.  Instead of showing full subtitles, our skimmed version can be used as an alternative text captioning support for TV broadcasting.

\subsection{Real-time Subtitling without scripts}

We also consider a real-time version, where no script is available as the source of subtitle.  In this case, the system solely relies on the speech recognition and keyword skimming is also incrementally applied.  We are planning to use the Google Cloud Speech API~\cite{googleSpeech} that provides a function where the recognized word sequence can be available before the pronounced sentence ends.  Such a real-time system should be useful as a wearable device that continuously supports the wearer's listening effort in daily life.

With such a real-time version, there might be a slight delay between text captioning and the word utterance, even with the real-time speech recognition. In this case, a listener might miss an important word and the subtitle of the corresponding word appears after some delay.  Thus the listener might have a cognitive load for integrating these two events. We are planning to investigate this issue more closely to reveal the effect caused by the delay.

\section{Conclusion}
We introduced \textit{Flash Word}, which is aim to improve ESL students' listening skills. With the prototype system we implemented we were unable to show that the system helped the participants to listen to English audio better than normal subtitles. However, it is suggested from the interviews that advanced ESL students appreciate \textit{Flash Word} better. In future we plan to show if there is any English level which benefits from the system

\bibliographystyle{ACM-Reference-Format}
\bibliography{sample}

\end{document}